\documentstyle[prl,aps,twocolumn]{revtex}
\begin{document}
\draft
\title{Shadow band in the one--dimensional large $U$ Hubbard model}
\author{
  Karlo Penc$^{a}$\cite{*}, Karen Hallberg$^{a}$, Fr\'ed\'eric Mila$^{b}$
  and Hiroyuki Shiba$^{c}$
}

\address{
     $(a)$ Max-Planck-Institut f\"ur Physik komplexer Systeme,
          Bayreuther Str.~40, 01187 Dresden (Germany) \\
     $(b)$ Laboratoire de Physique Quantique, Universit\'e Paul Sabatier,
     31062 Toulouse (France) \\
     $(c)$ Tokyo Institute of Technology, Department of Physics,
           Oh-okayama, Meguro-ku, Tokyo 152 (Japan)
}

\maketitle

\begin{abstract}
We show that the factorized wave--function of Ogata and Shiba can be used to
calculate the $k$ dependent spectral functions of the one--dimensional,
infinite
$U$ Hubbard model, and of some extensions to finite $U$. The resulting spectral
function is remarkably rich: In addition to low energy features typical of
Luttinger liquids, there is a well defined band, which we identify as the
shadow
band resulting from $2k_F$ spin fluctuations. This band should be detectable
experimentally because its intensity is
comparable to that of the main band for a large range of momenta.
\end{abstract}

\pacs{79.60.-i, 71.10.Fd, 78.20.Bh}

\narrowtext
The calculation of the spectral functions of models of correlated electrons
is one the
most challenging and largely unsolved issues of condensed matter theory.
Although a
number of numerical techniques can be used, e.g. exact
diagonalization of finite
clusters\cite{ed} or quantum Monte Carlo simulations\cite{montecarlo},
 exact results are available
only in very special cases, mostly for one--dimensional
spin models\cite{shsu}.
As far as one--dimensional electron
models are concerned, most of the well established results have been obtained
 in the
framework of the Luttinger liquid theory\cite{solyom,haldane,schulz,specfunc},
which is
believed to be the correct
description of the low energy properties of a large class of Hamiltonians.
 However, an accurate determination
of the dynamical properties for all frequencies is so far still
lacking.

In this paper we perform such a calculation for
the following one--dimensional models:

i) The Hubbard model defined by the Hamiltonian
\begin{equation}
  {\cal H}= -t \sum_{i,\sigma}
   \left(
    c^\dagger_{i,\sigma} c^{\phantom{\dagger}}_{i+1,\sigma}
   + h.c. \right)
   + U \sum_{i} n_{i,\uparrow} n_{i,\downarrow}
\end{equation}
in the infinite $U$ limit, which is also
    equivalent to the $J\to 0$ limit of the standard $t-J$ model;

ii) An extension of the $t-J$ model first proposed by Xiang and d'Ambrumenil
\cite{XA} defined by the Hamiltonian
\begin{eqnarray}
  {\cal H}&=& -t \sum_{i,\sigma}
   \left(
    \tilde c^\dagger_{i,\sigma} \tilde c^{\phantom{\dagger}}_{i+1,\sigma}
   + h.c. \right) \nonumber\\
   & & + \sum_{i,j} \sum_{\alpha=x,y,z}
     J^\alpha \left( S_i^\alpha S_{i+j}^\alpha
     - \case{1}{4} \delta_{\alpha,z} n_i n_{i+j}\right) {\cal P}_{i,j} \>,
   \label{eq:xiang}
\end{eqnarray}
where $\tilde c$ are the usual projected operators and
${\cal P}_{i,j} = \prod_{j'=1}^{j-1} (1 \! - \! n_{i+j'})$
in the exchange part of the Hamiltonian ensures that two spins
interact as long as there is no other spin between them.
The motivation to study this model is that, unlike the infinite $U$
Hubbard model, there is an
energy $J$ associated to spin fluctuations, and this will give us useful
indications
about the $1/U$ corrections in the case of the finite $U$ Hubbard model.

Although the Hamiltonians of the two models are different, they
share the remarkable property
that in both cases the eigenstates can be factorized\cite{XA,shiba,paso,XYtJ}
as
\begin{equation}
  |f,N \rangle = |\psi^{N}_{L}(Q,\{I\})\rangle
         \otimes |\chi_N(Q,\tilde f_Q) \rangle \>,
  \label{eq:OSwfLHB}
\end{equation}
where $|\psi^{N}_{L}(Q,\{I\})\rangle$ is an eigenfunction of
$N$ non--interacting spinless fermions on $L$ sites with momenta
$  k_j L = 2 \pi I_j + Q $ ($I_j$ are integers, $j=1\dots N$)
and $|\chi_N(Q,\tilde f_Q) \rangle$ is
an eigenfunction of the one dimensional spin--$\case{1}{2}$ Heisenberg model
with $N$ spins
(we choose $N$ as even integer not multiple of four)
and momentum $Q = 2 \pi J/N$, $J$ integer. This momentum imposes a twisted
boundary condition with phase $e^{iQ}$ to the spinless fermions\cite{sorella2}.
For more details, see Ref.~\onlinecite{local}.
This wave--function has already been used by Ogata and Shiba\cite{shiba}
to calculate the momentum distribution function and by Penc, Mila and Shiba to
calculate the local spectral function of the infinite $U$ Hubbard
model\cite{local}.

In the following, we will determine the full momentum dependence
of the photoemission and inverse photoemission spectral
functions defined by
\begin{eqnarray}
  A(k,\omega) &=&
 \sum_{f,\sigma}
  \left| \langle f,N \!+\! 1| c^\dagger_{k,\sigma} |0,N\rangle \right|^2
  \delta(\omega \!-\! E^{N+1}_f \!+\! E^N_0)
  \>,\nonumber\\
  B(k,\omega) &=&
 \sum_{f,\sigma}
  \left| \langle f,N \!-\! 1| c^{\phantom{\dagger}}_{k,\sigma} |0,N\rangle
\right|^2
    \delta(\omega \!-\! E^{N}_0 \!+\! E^{N-1}_f)
  \>.
\nonumber
\end{eqnarray}

As a result of the factorized form of the wave
functions, the
spectral functions can be obtained as a convolution:
\begin{eqnarray}
  A^{\rm LHB}(k,\omega)
  & = &
 \sum_{\omega',Q,\sigma}
  C_{\sigma}(Q,\omega')
  A_{Q}(k,\omega-\omega')
  \>, \nonumber\\
  B(k,\omega)
  &=&
 \sum_{\omega',Q,\sigma}
  D_{\sigma}(Q,\omega')
  B_{Q}(k,\omega-\omega')
  \>.
  \label{eq:blhbca}
\end{eqnarray}
A similar expression holds for the spectral function in the upper Hubbard
band\cite{longpaper}
 $A^{\rm UHB}(k,\omega\approx U)$, which we will not discuss here.
In these expressions, $A_{Q}(k,\omega)$ and
$B_{Q}(k,\omega)$ involve only the spinless fermion part of the wave
 function and are defined as
\begin{eqnarray}
  A_{Q}(k,\omega)
 & = & L
 \sum_{\{I\}}
  \left|
    \langle \psi_{L}^{N+1}(Q,\{I\}) |
    b^\dagger_{0}
    | \psi_{L,\pi}^{N,{\rm GS}} \rangle
    \right|^2 \nonumber\\
   &&
   \times
  \delta(\omega \!-\! E^{N+1}_f \!+\! E^N_0)
  \delta(k\!-\! P^{N+1}_f \!+\! P^{N}_0 )
  \>,\nonumber\\
  B_{Q}(k,\omega)
 & = & L
 \sum_{\{I\}}
  \left|
    \langle \psi_{L}^{N-1}(Q,\{I\}) |
    b^{\phantom{\dagger}}_{0}
    | \psi_{L,\pi}^{N,{\rm GS}} \rangle
    \right|^2\nonumber\\
   &&\times
  \delta(\omega \!-\! E^N_0 \!+\! E^{N-1}_f )
  \delta(k\!-\! P^{N-1}_f \!+\! P^{N}_0)
 \>,\label{eq:defaomegaQ}
\end{eqnarray}
where the momentum and energy of the states are given by
$P^{N'} = \sum_{j=1}^{N'} k_j$ and $E^{N'} = -2 t \sum_{j=1}^{N'} \cos k_j$,
and where $b$
and $b^\dagger$ are spinless fermion operators.
$C_{\sigma}(Q,\omega)$ and  $D_{\sigma}(Q,\omega)$ depend
on the spin wave function only and are given by
\begin{eqnarray}
  C_{\sigma}(Q,\omega) &=& \sum_{\tilde f_Q}
    \left| \langle \chi_{N+1} (Q,\tilde f_Q) |
      \hat Z^\dagger_{0,\sigma}
      | \chi_N^{\rm GS} \rangle
    \right|^2 \nonumber \\
  && \times \delta(\omega - E^{N+1}_f + E^{N}_0 )
  \>,\nonumber \\
  D_{\sigma}(Q,\omega) &=& \sum_{\tilde f_Q}
    \left| \langle \chi_{N-1}(Q,\tilde f_Q) |
      \hat Z_{0,\sigma}
      | \chi_{N}^{\rm GS} \rangle
    \right|^2 \nonumber \\
  && \times \delta(\omega - E^{N}_0 + E^{N-1}_f )
  \>,\label{eq:defcomegaQ}
\end{eqnarray}
where
$\hat Z^{\dagger}_{0,\sigma}$
appends a spin $\sigma$ to the beginning of the spin wave function
$|\chi_N \rangle$ making it $N+1$ sites long, and  $\hat Z_{0,\sigma}$ is the
hermitian conjugate of $\hat Z^{\dagger}_{0,\sigma}$.

To evaluate the charge contribution, one needs matrix elements
between states with different boundary conditions
($e^{iQ}$ for the final state,
$e^{i\pi}$ for the ground state\cite{local}).
For  $Q \neq \pi$  the overall phase shift $(Q-\pi)/L$ due to momentum
transfer $Q-\pi$ to the spin degrees of freedom
gives rise to
Anderson's orthogonality catastrophe\cite{ortho1} and the matrix
elements
$ |
   \langle
     \psi_{N+1}(Q,\{I\}) | b^\dagger_{0}| \psi_{N}^{\rm GS}
   \rangle
 |^2$ can be shown to be equal to
\begin{eqnarray}
 && L^{-2N-1}
  \cos^{2N}\frac{Q}{2}
   \prod_{j>i} \sin^2 \frac{k_j-k_i}{2} \nonumber\\
 &&\times  \prod_{j>i} \sin^2 \frac{k'_j-k'_i}{2}
   \prod_{i,j} \sin^{-2} \frac{k'_i-k_j}{2}
   \>,
\end{eqnarray}
where $k_j$ ($k'_j$) are wave vectors with phase shift $Q/L$ ($\pi/L$). The
restriction imposed by $\delta(k- P^{N\pm 1}_f + P^{N}_0 )$ is then implemented
by restricting the sum over $\{I\}$ to states which have the correct momentum.

The calculation of the spin contribution is based on the
spin--$\case{1}{2}$ Heisenberg Hamiltonian with $N'$ sites
\begin{equation}
  {\cal H}_{\rm spin} =  \sum_{i=1}^{N'} \sum_{\alpha=x,y,z}
    \tilde J^\alpha
  \left(S_i^\alpha S_{i+1}^\alpha -\case{1}{4} \delta_{\alpha,z}\right)
   \>,
\end{equation}
with $N'=N$ for the ground--state and $N\pm 1$ for the final states.
The model of Eq.~(\ref{eq:xiang}) corresponds to
$\tilde J^\alpha = J^\alpha$. For the infinite $U$ Hubbard model, one has to
consider the isotropic case and to take
the limit $\tilde J \rightarrow 0$. In that case,
there is no energy associated
to spin
excitations, and we can write
$C_{\sigma}(Q,\omega) = C_{\sigma}(Q) \delta(\omega)$ and
$D_{\sigma}(Q,\omega) = D_{\sigma}(Q) \delta(\omega)$.
The functions $C_{\sigma}(Q) $ and $D_{\sigma}(Q) $ have already been
studied in our previous paper\cite{local}.
They can be calculated numerically with exact
digonalizations (up to 26 sites) or with DMRG\cite{white}
 (up to 130 sites, keeping 300 states per block).
It turns out that there is a very strong singularity $(\pi/2-Q)^{-1/2}$
for $Q<\pi/2$ and some background coming from the higher order excitation
towers
for $Q>\pi/2$ in the case of $D_{\sigma}(Q,\omega)$. For $C_{\sigma}(Q,\omega)$
the situation is reversed and both are symmetric with respect to $Q=0$.

Using these results, it is straightforward to get the spectral
functions for the
infinite $U$ Hubbard model. One just has to
generate the quantum numbers for the charge part, calculate the corresponding
energy, momentum and matrix elements, and perform the convolution in $Q$.
The results are presented in
Fig.~\ref{fig:3d} and ~\ref{fig:2dH} for a quarter--filled system.
There are several interesting features
to notice. In the low energy region near $k_F$ we can identify three
structures. For $k<k_F$ there are divergences
at $\omega=u_c (k-k_F)$ and $\omega=0$
and a lot of spectral weight
between them (peaks `b' and `c' on Fig.~\ref{fig:2dH} ).
There is also a small weight (`e')
appearing on the other side of the Fermi energy for $\omega>-u_c(k-k_F)$.
For $k>k_F$
the spectrum is symmetric with respect to $k_F$. If we remember
that the spin velocity $u_s$ vanishes for the infinite $U$ Hubbard model, all
these features are consistent with the Luttinger liquid calculations of
Meden and Sch\"onhammer\cite{specfunc} and of
Voit\cite{specfunc}.
The small peak `g' comes from higher harmonics.
The dispersion of the charge part (`b') is exactly given
by $E(k)=-2t \cos(|k|+k_F)$, in agreement with the observation of Preuss et
al\cite{montecarlo}
based on Monte Carlo results for $U/t=4$.

However, the Luttinger liquid picture does not exhaust the features of the
spectral function of Fig.~1 and 2. For larger energies, or away from
$k_F$,
there is a well defined band--like structure (`a') with
considerable spectral weight and a dispersion given by
$E(k)=-2t \cos(-|k|+k_F)$.
We interpret this feature as a shadow band\cite{shadow} coming
from the spin fluctuations  which diverge at $2k_F$. The scattering of
the charges by these fluctuations produces an image of the main spectrum
at $k+2 k_F$. This is very similar to the mechanism of the shadow bands
proposed
for the two dimensional model with strong antiferromagnetic fluctuations.
This shadow band
is responsible for the singularity at $3k_F$ present in the momentum
distribution function\cite{shiba,Greensfunc3kF,ren}.
Finally, there is a Van Hove singularity
at $\pm 2t$ which gives rise to a clear peak for wave vectors close to the
extrema of the bands (`f').

Let us now turn to the model of Eq.~(\ref{eq:xiang}).
To get the spectral function, we need
$C_{\sigma}(Q,\omega)$ and $D_{\sigma}(Q,\omega)$ for the Heisenberg
model. This can be done numerically for the isotropic case ($J^{x,y,z} = J$)
using
L\'anczos diagonalization of small clusters or DMRG\cite{karendyn}.
We find
that $C_{\sigma}(Q,\omega)$ is zero for $\omega< - \tilde J \ln 2
+ u_\sigma |\sin(k-\case{\pi}{2})|$, where $u_\sigma=\case{\pi}{2}\tilde J$ is
the spin velocity in the squeezed system,
that it has an inverse square root singularity at $Q=\pi/2$, and that
the largest
contributions come from the lower edge of the excitation spectrum.
The main difference with the infinite $U$ case is that the spin fluctuations
have an energy of order $J$, so that the spin velocity $u_s = u_\sigma L/N$
does not vanish anymore. The low energy part of the spectrum has then
exactly the form predicted by the Luttinger liquid theory.

For the $XY$ case ($J^{x,y} = J$, $J^z = 0 $)
one can give a closed expression for $C_{\sigma}(Q,\omega)$ and
$D_{\sigma}(Q,\omega)$ after mapping the problem onto non--interacting
spinless fermions by a Jordan--Wigner transformation. After some algebra,
the matrix elements
$|\langle \chi_{N+1} (Q,\tilde f_Q) |Z^\dagger_{0,\sigma}|
 \chi_N^{\rm GS} \rangle|^2 $
of Eq.~(\ref{eq:defcomegaQ}) can be obtained as
\begin{eqnarray}
  &&[N(N+1)]^{-M}
  \prod_{j=1}^M \sin^2 \frac{q'_j}{2}
   \prod_{j>i} \sin^2 \frac{q_j-q_i}{2}
  \nonumber \\
  &&\times
   \prod_{j>i} \sin^2 \frac{q'_j-q'_i}{2}
   \prod_{i,j} \sin^{-2} \frac{q'_i-q_j}{2}
   \>,
\end{eqnarray}
where $q_j$ and $q'_j$ are the momenta of the $M$ spinless fermions
representing
the $-\sigma$ spins on the $N$ and $N+1$ site lattice. They are
quantized according to $q'=2\pi J'_j/(N+1)$ and  $q=2\pi J_j/N$, where
$J_j$ and $J'_j$ are integer quantum numbers,
and $\tilde f_Q \equiv \{J'_j,j=1..M\}$.
The total momentum and energy of $|\chi_{N+1}(Q,\tilde f_Q)\rangle$ are
given by $Q=\sum_j q'_j$ and $E^{N+1}= J \sum_j \cos q'_j$. Details will be
given elsewhere \cite{longpaper}.
A similar expression holds in the case of
$D_\sigma(Q,\omega)$.
This formulation also allows one to derive analytical results. For instance,
the static function $\omega(0 \rightarrow j,\sigma)$ introduced by Ogata
and Shiba \cite{shiba} can be shown to have the asymptotic behavior
$\propto j^{-5/8} \cos( \case{\pi}{2}j + \case{\pi}{4})$.
Thanks to this mapping, one can calculate the spectral function
with the same accuracy as for the infinite $U$ Hubbard model. The results are
shown in
Fig.~\ref{fig:XYcut} for a quarter--filled system.
It is essentially the same as that of the Hubbard model,
except that at low energies an extra peak `d' accounting for the extra
exponents in the spin part of the XY model has appeared
(this peak has nothing to do with
peak `g' on Fig.~2.) Due to finite $J$, both `c' and `d'
follow the $\omega = u_\sigma \cos\case{\pi}{2}\case{k}{k_F}$ dispersion.
Furthermore, we can see that the shadow band (`a') and the Van Hove like
singularity (`f') are broadened by the spin fluctuations.

Finally, let us comment on the experimental implications of the
present results.
It would be most interesting to observe the shadow band in angular-resolved
photoemission or inverse photoemission experiments on quasi-one dimensional
conductors. The intensity of that band in the previous calculations is
 certainly big enough for it to be detected.
What about the experimentally more relevant case of the Hubbard model
with finite $U$? In that case the
factorized wave functions are no longer eigenfunctions of the Hubbard
model, and there are two types of $1/U$ corrections to the spectral
functions.
The first type is due to the energy
coming from the spin part with an effective coupling
$\tilde J \approx \case{4 t^2}{U} (n-\case{\sin 2 \pi n}{2\pi})$.
We expect these corrections to be very similar to those of the
model of Eq. (\ref{eq:xiang}), and the main effect is to give a finite velocity
to the spin excitations.
However, there are also $1/U$ corrections entering the matrix elements of the
spinless
part of the wave--function.
We can anticipate that they will have two effects on the spectral function.
They will produce a transfer of spectral weight to the upper Hubbard
band which, according to Eskes and Ole\'s\cite{oles}, will be small except very
close to half--filling, and they will modify the power laws of the
singularities. So, at least not too close to half--filling, the shadow band
seems to be robust against $1/U$ corrections.
Whether this remains true for
small values of $U$ is not clear yet.
Let us just mention that, according to recent
numerical results  obtained  by Maekawa et al.\cite{maekawa}
in a study of the spectral function of the Hubbard model for $U/t=10$
based on L\'anczos diagonalization of finite clusters,
there seems to be a structure in addition to the Luttinger liquid
features, suggesting
that $U/t=10$ is already large enough to guarantee
the presence of a well defined shadow band.

We thank to Y. Kuramoto, H. Fukuyama, M. Imada, D. Poilblanc, K. Vlad\'ar and
A. Zawadowski
for useful discussions.

\begin{figure}
\caption{
One particle spectral functions of the $U\rightarrow +\infty$ Hubbard model for
$L=228$ sites and $N=114$ electrons with Fermi momentum $k_F=\pi/4$.}
\label{fig:3d}
\end{figure}

\begin{figure}
\caption{The same as Fig.~1, but for some selected momenta. Some
parts of the spectra are
multiplied by 10 and are shown with dashed lines.}
\label{fig:2dH}
\end{figure}

\begin{figure}
\caption{Spectral function for the model of Xiang and d'Ambrumenil
 with XY exchange,
$J=0.4 t$, $L=228$, $N=114$ and $\varepsilon_F = -J /\pi$.
Some parts of the spectra are
multiplied by 10 and are shown with dashed lines.}
\label{fig:XYcut}
\end{figure}


\begin{references}

\bibitem[*]{*}
  On leave from Research Institute for Solid State Physics, Budapest, Hungary.

\bibitem{ed} See e.g. E. Dagotto, Rev. Mod. Phys. {\bf 66}, 763 (1994).

\bibitem{montecarlo}
  R. Preuss {\it et al.}, Phys. Rev. Lett. {\bf 73}, 732 (1994).

\bibitem{shsu}
G. M\"uller and R. E. Shrock,  Phys. Rev. Lett. {\bf 51}, 219 (1983) and
references therein;
Z. N. C. Ha and F. D. M. Haldane, Phys. Rev. Lett. {\bf 73}, 2887 (1994);
B. D. Simons, P. A. Lee and B. I. Altshuler, {\it ibid} {\bf 70}, 4122 (1993);
F. D. M. Haldane and M. R. Zirnbauer, {\it ibid} {\bf 71}, 4055 (1993).

\bibitem{solyom}
 J. S\'olyom, Adv. Phys. {\bf 28}, 201 (1979).

\bibitem{haldane}
  F. D. M. Haldane, J. Phys. C {\bf 14}, 2585 (1981)

\bibitem{schulz}
  H. J. Schulz, Phys. Rev. Lett. {\bf 64}, 2831 (1990); Int. J. Mod. Phys.B
{\bf
  5}, 57 (1991).

\bibitem{specfunc}
  V. Meden and K. Sch\"onhammer, Phys. Rev. B {\bf 46}, 15753 (1992);
  K. Sch\"onhammer and V. Meden, ibid. {\bf 47}, 16205 (1993);
  J. Voit, ibid.  {\bf 47}, 6740 (1993).

\bibitem{XA}
  T. Xiang and N. d`Ambrumenil, Phys. Rev. B {\bf 45}, 8150 (1992).

\bibitem{shiba}
  M. Ogata, T. Sugiyama and H. Shiba, Phys. Rev. B {\bf 43}, 8401 (1991);
  M. Ogata and  H. Shiba, ibid {\bf 41}, 2326 (1990).

\bibitem{paso}
  A. Parola and S. Sorella, Phys. Rev. Lett. {\bf 64}, 1831 (1990).

\bibitem{XYtJ}
  T. Prushke and H. Shiba, Phys. Rev. B {\bf 44}, 205 (1991).

\bibitem{sorella2}
  S. Sorella and A. Parola, J. Phys. Condens. Matter {\bf 4}, 3589 (1992).

\bibitem{local}
 K. Penc, F. Mila and H. Shiba, Phys. Rev. Lett. {\bf 75}, 894 (1995).

\bibitem{longpaper}
  K. Penc, K. Hallberg, F. Mila, H. Shiba, unpublished.

\bibitem{ortho1}
  P. W. Anderson, Phys. Rev. Lett. {\bf 18}, 1049 (1967);
  G. Yuval and P. W. Anderson, Phys. Rev B {\bf 1}, 1522 (1970);

\bibitem{white} S. R. White, Phys. Rev. Lett. {\bf 69}, 2863 (1992).

\bibitem{shadow} A. P. Kampf and J. R. Schrieffer, Phys. Rev. B {\bf 42}, 7967
(1990). For recent developments in 2D, see S. Haas et al, Phys. Rev. Lett. {\bf
74}, 310 (1995); R. Preuss et al. Phys. Rev. Lett. {\bf 75}, 1344 (1995); A.
Chubukov, Phys. Rev. B {\bf 52}, R3840 (1995).

\bibitem{Greensfunc3kF}
  K. Penc and J. S\'olyom, Phys.~Rev.~B{\bf 44}, 12690 (1991)

\bibitem{ren}
  P. W. Anderson and Y. Ren , in High Temperature Superconductivity, edited by
  K. S. Bedell {\it et al.} (Addison Wesley, Redwood City, 1990), p. 3.

\bibitem{karendyn} K. Hallberg, Phys. Rev. B {\bf 52}, R9827 (1995).

\bibitem{oles}
  H. Eskes and A. M. Ole\'s, Phys. Rev. Lett. {\bf 73}, 1279 (1994);


\bibitem{maekawa} S. Maekawa, T. Tohyama and S. Yunoki, unpublished.

\end{references}
\end{document}